\documentclass[journal]{IEEEtran}

\usepackage{cite}

\usepackage[bookmarks, colorlinks=true, plainpages = false, citecolor = blue,linkcolor=red,urlcolor = blue, filecolor = blue,pagebackref]{hyperref}

\usepackage{setspace,multicol,color}

\usepackage{graphicx,graphics}
\usepackage{subfig,epsfig}

\usepackage{bm,amsmath,amssymb,amsthm,enumerate}

\usepackage{algorithmicx}
\usepackage[linesnumberedhidden,ruled,vlined]{algorithm2e}

\usepackage{wrapfig}

\theoremstyle{plain}

\theoremstyle{definition}
\theoremstyle{remark}

\newcommand{\bLambda}{{\bm{\Lambda}}}

\newcommand{\bSigma}{{\bm{\Sigma}}}

\newcommand{\cA}{{\mathcal{A}}}

\newcommand{\cF}{{\mathcal{F}}}

\newcommand{\cS}{{\mathcal{S}}}

\newcommand{\cH}{{\mathcal{H}}}

\newcommand{\cN}{{\mathcal{N}}}

\newcommand{\ba}{{\boldsymbol a}}
\newcommand{\bb}{{\boldsymbol b}}

\newcommand{\bu}{{\boldsymbol u}}

\newcommand{\bw}{{\boldsymbol w}}
\newcommand{\bx}{{\boldsymbol x}}
\newcommand{\by}{{\boldsymbol y}}

\newcommand{\E}{{\mathbb E}}
\newcommand{\R}{{\mathbb R}}
\newcommand{\C}{{\mathbb C}}

\newcommand{\bA}{{\boldsymbol A}}
\newcommand{\bB}{{\boldsymbol B}}
\newcommand{\bC}{{\boldsymbol C}}
\newcommand{\bD}{{\boldsymbol D}}

\newcommand{\bI}{{\boldsymbol I}}

\newcommand{\bK}{{\boldsymbol K}}
\newcommand{\bL}{{\boldsymbol L}}

\newcommand{\bP}{{\boldsymbol P}}
\newcommand{\bQ}{{\boldsymbol Q}}
\newcommand{\bR}{{\boldsymbol R}}

\newcommand{\bU}{{\boldsymbol U}}
\newcommand{\bV}{{\boldsymbol V}}

\newcommand{\bX}{{\boldsymbol X}}

\newcommand{\Tr}{\mathop{\rm Tr}}

\newcommand{\diag}{\mathop{\rm diag}}

\newcommand{\vect}{\mathop{\rm vec}}
\newcommand{\grad}{{\triangledown}}

\renewcommand{\Re}{\operatorname{Re}}

\begin{document}
%
\title{Low-rank LQR Optimal Control Design over Wireless Communication Networks} 
%
%
%


\author{Myung~Cho, Abdallah~Abdallah, and Mohammad~Rasouli
\thanks{M. Cho, A. Abdallah, M. Rasouli are with the Department of Electrical and Computer Engineering, Behrend College, Penn State University, Erie, PA, 16563, USA
        {\tt\small (email: (mxc6077,aua639,mur37)@psu.edu)}}
}

%
%

\markboth{}%
{Shell \MakeLowercase{\textit{et al.}}: Bare Demo of IEEEtran.cls for IEEE Journals}
%



\maketitle

\begin{abstract}
This paper considers a LQR optimal control design problem for distributed control systems with multi-agents. To control large-scale distributed systems such as smart-grid and multi-agent robotic systems over wireless communication networks, it is desired to design a feedback controller by considering various constraints on communication such as limited power, limited energy, or limited communication bandwidth, etc. In this paper, we focus on the reduction of communication energy in an LQR optimal control design problem on wireless communication networks. By considering the characteristic of  wireless communication, i.e., Radio Frequency (RF) signal can spread in all directions in a broadcast way, we formulate a low-rank LQR optimal control model to reduce the communication energy in the distributed feedback control system. To solve the problem, we propose an Alternating Direction Method of Multipliers (ADMM) based algorithm. Through various numerical experiments, we demonstrate that a feedback controller designed using low-rank structure can outperform the previous work on sparse LQR optimal control design, which focuses on reducing the number of communication links in a network, in terms of energy consumption, system stability margin against noise and error in communication. 
\end{abstract}


%
\IEEEpeerreviewmaketitle

\section{Introduction}
\label{sec:intro}
Design of feedback control systems has been studied for several decades and applied to various applications in autonomous vehicles, power plants, and robots to name a few. Optimal control design methods can be used optimizing various objective functions (e.g., Linear Quadratic Regulator (LQR), $\cH_2$ norm or $\cH_{\infty}$ norm) to meet some design criteria in a feedback controller. 

Unlike optimal control design for conventional systems that are studied in \cite{rautert1997computational,zhou1996robust,peres1994alternate} and references therein, recent control systems can be very different from the previous ones in various aspects. Recent systems are larger in scale, distributed, ubiquitous, and connected via wireless communication network. The evolution of wireless communication devices such as cellular and Internet of Things (IoTs) devices significantly contribute to recent changes in control systems. Thus, recent control systems may have multi-agents distributed in large-scale topology, which communicate over wireless communication networks. With the new paradigm of distributed systems, we face new challenges including the communication energy overhead, response time and delay, privacy and security issues, etc. 

To address the new challenges and issues in distributed multi-agents control systems, especially reducing communication burden, several research studies have been conducted. The main focus has been on the network connectivity aspect of designing a feedback control system. More specifically, in \cite{fazelnia2016convex,bamieh2002distributed,de2002convex,d2003distributed,bamieh2005convex,motee2008optimal,fardad2009optimal,fardad2014design,lin2011augmented,fardad2011design} and the references therein, LQR control designs with predetermined structure of network topologies were studied. In order to reduce the number of communication links in a distributed multi-agents system, the proposed solutions in \cite{lin2013design,fardad2009optimal,lian2017game,cho2022iterative,cho2020communication} took into account sparse LQR control design models which simultaneously minimize LQR cost as well as sparsity level of the network topology by considering the sparsity condition on the topology as a regularization term or as a constraint in optimization problems. 

The research studies conducted so far raise the following research question: ``Is reducing the number of communication links among agents helping to reduce the total energy consumed in both control and communication operations?'' For example, with the reduced number of communication links, we may reduce the communication energy, but, what if we need to spend more energy, e.g., LQR cost, in control? Then, reducing the communication links may not result in reducing the total energy spent in the whole process. In this paper, we attempt to answer this question in a distributed control system with multi-agents connected via a wireless communication network. The idea is that when wireless nodes run in a broadcast mode, the increase in network coverage may help to reduce communication energy and delay with minimal impact on the LQR cost, compared to the standard LQR optimal control design and sparse LQR control design. Therefore, we formulate low-rank LQR optimal control design problems, and propose algorithms to solve the problems.

The contribution of this paper is three-fold. First, we introduce new LQR optimal control problems, which we call ``low-rank LQR'' control design problems. The sparse LQR control design applies sparseness to the structure of a feedback matrix, while in the low-rank LQR control design problems, we consider low-rankness on the feedback matrix, which can be interpreted as a controller with communication in a broadcast mode. Secondly, to solve these novel optimization problems, we introduce Alternating Direction Method of Multipliers (ADMM) algorithms. Finally, we demonstrate that under various wireless communication scenarios, our proposed method outperforms the previous solutions that utilize standard LQR control and sparse LQR control designs.

The rest of the paper is organized as follows. Section \ref{sec:signal} introduces the problem statement for the optimal control design that minimizes the LQR cost with a low-rank constraint on communication networks, and describes how this structure of a wireless network is interpreted in control. In Section \ref{sec:review}, we briefly review the previous research on standard LQR and sparse LQR control designs. In Section \ref{sec:algo}, we describe the merit of the low-rank LQR control design against the standard LQR and the sparse LQR control designs, and propose the ADMM based algorithm to solve the low-rank LQR optimal control problems. In Section \ref{sec:experiment}, we provide numerical experiment results demonstrating the performance of our proposed work against the standard LQR control and the sparse LQR control designs under various communication scenarios. Finally, Section \ref{sec:conclusion} concludes the paper and introduce possible future research directions.

\textbf{Notations}: $\R$  and $\C$ are reserved for the sets of real numbers and complex numbers respectively. We denote a scalar, a vector, and a matrix as a non-bold letter, a bold small letter, and a bold capital letter respectively, e.g., $x$ or $X$ for a scalar, $\bx$ for a vector, $\bX$ for a matrix. We denote $\Re(\cdot)$ as the real part of a complex value. We use the super-script $T$ for transpose. For a matrix $\bA \in \R^{m \times n}$, we use Frobenius norm as $||\bA||_F$,  element-wise $\ell_1$ norm as $\|\bA\|_1$, i.e., $\|\bA\|_1=\sum_{i,j} |A_{i,j}|$, and nuclear norm as $\| \bA \|_{*}$, i.e., sum of singular values of $\bA$, respectively. We reserve $\bI$ for the identity matrix. A feasible set of feedback matrices $\bK$'s with asymptotic stability is denoted as $\cF$, i.e., $\cF := \{  \bK \;|\;  \max( \Re( \lambda ( \bA - \bB_1\bK ) ) ) < 0 \}$, where $\lambda(\bA - \bB_1 \bK)$ represents the eigenvalue of the closed-loop state matrix $\bA - \bB_1 \bK$. For a matrix $\bQ$, $\bQ \succeq 0$ and $\bQ \succ 0$ represent a symmetric positive semidefinite matrix and a symmetric positive definite matrix respectively. $\otimes$ represents the Kronecker product.

\section{Problem Statement}
\label{sec:signal}
In this paper, we consider a distributed control system with multiple agents where we need a communication network to share feedback signals among the agents to stabilize the whole system in a feedback loop. This system can be expressed as the following state space representation:
\par
\begin{figure}[t]
    \centering
    \subfloat[Illustration of a distributed system with four agents]{\includegraphics[scale=0.65]{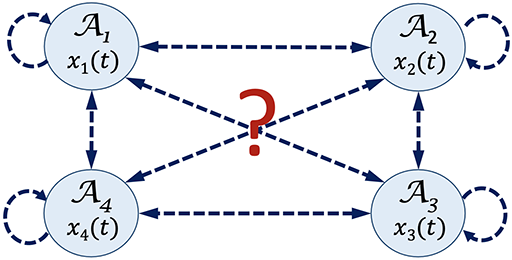}}\\
    \subfloat[Corresponding feedback signal]{\includegraphics[scale=0.35]{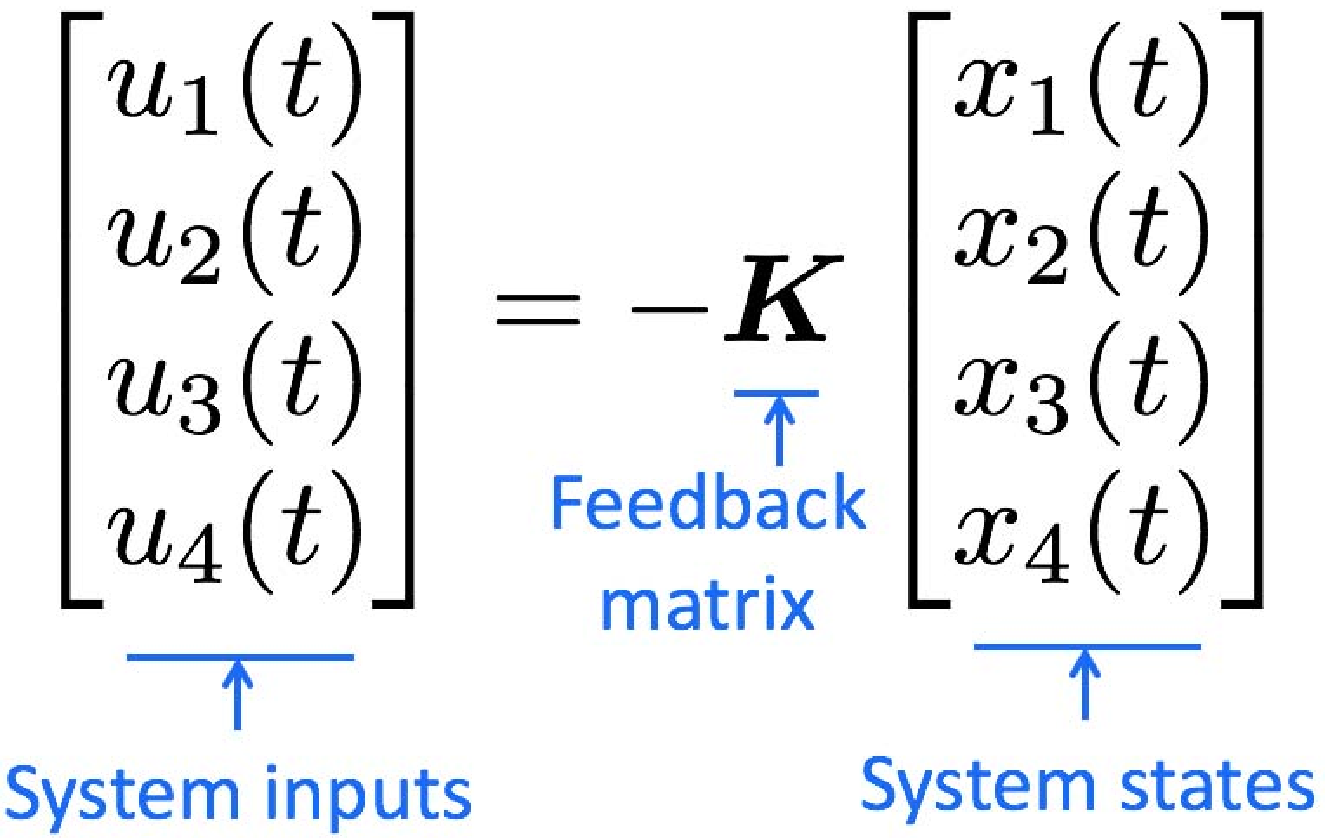}}
    \caption{Illustration of a distributed system with four agents denoted by $\cA_1$, $\cA_2$, $\cA_3$ and $\cA_4$, where a dotted arrow represents a feedback signal from one agent to another or an internal feedback signal. In (b), $\bu_i(t)$ and $\bx_i(t)$ represent input and state vectors of the $i$-th agent, and each agent has its derivative $\dot{\bx}_i(t)$ and integration part inside, which are omitted in (a).}
    \label{fig:DS}
\end{figure}

\noindent\small
\begin{align}\label{eq:SSE}
	& \dot{\bx}(t) = \bA \bx(t) + \bB_1 \bu(t) + \bB_2 \bw(t), \;(\bx(0)=\bB_2 \bw(0)),\nonumber  \\
	& \by(t) = \bC \bx(t) + \bD \bu(t), \nonumber \\
	& \bu(t) = -\bK \bx(t),
\end{align}
\normalsize
where $\bx(t)$, $\dot{\bx}(t)$, and $\bu(t)$ are the state vector, its derivative with respect to time, and the input vector respectively. We organize the system state $\bx(t)$ (resp. its derivative) by stacking the states (resp. its derivatives) of each agent in a vector as shown in Fig. \ref{fig:DS}(b).  The system input $\bu(t)$ at time $t$ is also organized by stacking the inputs of agents in the system as shown in Fig. \ref{fig:DS}(b). $\bw(t)$ is disturbance at time $t$ with  i.i.d. Gaussian distribution $\cN(0,\bI)$. Also, $\by(t)$ is the output of the system at time $t$. Correspondingly, a state matrix, input matrix, and disturbance are denoted by $\bA \in \R^{n \times n}$, $\bB_1 \in \R^{n \times m}$ and $\bB_2 \in \R^{n \times l}$ respectively. $\bC$ and $\bD$ are output matrices. $\bK \in \R^{m \times n}$ represents a feedback matrix. Throughout the paper, we assume that $(\bA, \bB_1)$ is stabilizable and $(\bA, \bQ^{1/2})$ is detectable. The goal here is to find a feedback matrix $\bK$ that makes not only the whole system asymptotically stable but also satisfies certain conditions, e.g., low LQR cost, low-rank, and sparsity, etc.

The state space representation \eqref{eq:SSE} can be restated as 
\par\noindent\small
\begin{align}\label{eq:SSE_feedback}
	& \dot{\bx}(t) = (\bA - \bB_1\bK)\bx(t) + \bB_2 \bw(t), \;(\bx(0)=\bB_2 \bw(0)), \nonumber  \\
	& \by(t) = (\bC - \bD\bK) \bx(t).
\end{align}
\normalsize
Due to the way how we organize the system input $\bu(t)$ and the system state $\bx(t)$ as shown in Fig. \ref{fig:DS}(b), non-zero entries in off-diagonal (or off-block-diagonal) of the feedback matrix $\bK$ pertain to communication links, i.e., dotted arrows in Fig. \ref{fig:DS}(a), among agents in the distributed system. For instance, if $u_i(t)$ and $x_i(t)$ are single-output functions over $t$, for all $i$, and we have non-zero entry in the $i$-th row and the $j$-th column of the feedback matrix $\bK$, then, there needs to be communication  from agent $\cA_j$ to agent $\cA_i$. Basically, the structure of the feedback matrix $\bK$ can be related to the communication links in a distributed system with multi-agents.

From \eqref{eq:SSE_feedback}, the LQR cost in infinite time domain is defined as follows:
\par\noindent\small
\begin{align}\label{def_J0}
J_0(\bK)  &  := \int_{t=0}^{\infty} \bx(t)^T \bQ \bx(t) + \bu(t)^T \bR \bu(t) \; dt \nonumber \\
			   & = \int_{t=0}^{\infty} \bx(t)^T ( \bQ + \bK^T \bR \bK) \bx(t) \; dt 
\end{align}
\normalsize
where $\bQ \succeq 0 \in \R^{n \times n}$ and $\bR \succ 0 \in \R^{m \times m}$ are given performance weight matrices. If $\bQ$ and $\bR$ are identity matrices, then, the LQR cost can be simply understood as the energy of a control system expected to be consumed in infinite time. Remark that the squared term on $\bx(t)$ can be related to the power of a signal $\bx(t)$, and the integral of the power over time can be interpreted as energy-related cost.

With the introduction of a matrix $\bP \in \R^{n \times n}$ such that $\frac{d}{dt} \bx(t)^T \bP \bx(t) = - \bx(t)^T (\bQ + \bK^T \bR \bK) \bx(t)$, we can express the expectation of $J_0(\bK)$, denoted by $J(\bK)$, over the disturbance as follows:
\par\noindent\small
\begin{align}
 J(\bK) & := \E \bigg[ \int_{t=0}^{\infty} -\frac{d}{dt} \bx(t)^T \bP \bx(t) dt \bigg] = \E\bigg[ \bx(0)^T \bP \bx(0)\bigg] \nonumber \\
 & \;= \Tr(\bB_2^T \bP\bB_2),
\end{align} 
\normalsize
where $\lim_{t \rightarrow \infty} \bx(t) = \bm{0}$ by the assumption of asymptotically stable feedback system, and the final equality is obtained from that $\bx(0) = \bB_2 \bw(0)$ and $\E [\bw(0) \bw(0)^T ] = \bI$.

Since we have $\frac{d}{dt} \bx(t)^T \bP \bx(t) = \dot{\bx}(t)^T \bP \bx(t) + \bx(t)^T \bP \dot{\bx}(t)$, where $\dot{\bx}(t) = (\bA - \bB_1\bK)\bx(t) + \bB_2 \bw(t)$ in \eqref{eq:SSE_feedback} and $\E[\bw(t)] = \bm{0}$, we have the following well-known Lyapunov equation over $\bK \in \R^{m \times n}$ and $\bP \in \R^{n \times n}$:
\par\noindent\small
\begin{align} \label{eq:Lyap_P}
	& ( \bA - \bB_1 \bK)^T \bP + \bP ( \bA - \bB_1 \bK) + \bQ + \bK^T \bR \bK = \bm{0},
\end{align}
\normalsize
where $\bP$ needs to be strictly positive definite. This Lyapunov equation can also be restated as $(\bI \otimes (\bA - \bB_1 \bK)^T + (\bA -\bB_1 \bK)^T \otimes \bI) \vect(\bP) = - \vect(\bQ +\bK^T \bR \bK)$, where $\vect(\cdot)$ is the vectorization operator for a matrix by stacking the columns of a matrix. From this equation, it is also recognized that if the feedback matrix $\bK$ is in $\cF$, then, all eigenvalues of the feedback system matrix $\bA - \bB_1 \bK$ have negative real parts.  Then, there will be no zero in the sum of any two eigenvalues of the feedback system matrix, which leads to $(\bI \otimes (\bA - \bB_1 \bK)^T + (\bA -\bB_1 \bK)^T \otimes \bI)$ is non-singular \cite{chen1998linear}. It indicates that for a given feedback matrix $\bK \in \cF$, there exists a unique matrix $\bP$. Hence, we consider $\bP$ as a function of $\bK$, denoted by $\bP(\bK)$.

Then, we introduce the LQR minimization problem with a regularization term for $\bK$ as follows:
\par\noindent\small
\begin{align}\label{eq:LQR_prob}
	& \underset{\bK}{\text{minimize}} \; J(\bK) + \gamma G(\bK) \;\nonumber \\
	& \text{subject to} \;   \bK \in \cF,
\end{align}
\normalsize
where $J(\bK) =\Tr(\bB_2^T \bP(\bK)\bB_2)$, $\bP(\bK)$ needs to satisfy the Lyapunov equation \eqref{eq:Lyap_P},  and $G(\bK)$ is a regularization term for the structure of the feedback matrix $\bK$, $\gamma \geq 0 $ is a tuning parameter for weighting the regularization term. Since the off-diagonal (or off-block-diagonal) of the feedback matrix $\bK$ is related to the communication links, let us decompose the feedback matrix $\bK$ into the sum of a diagonal matrix and a low-rank matrix as follows: 
\par\noindent\small
\begin{align}
	\bK = \bK_{diag} + \bK_{low}.
\end{align}
\normalsize
Then, we propose the low-rank LQR control design problem as follows:
\par\noindent\small
\begin{align}\label{eq:LQR_prob2}
	&\underset{\bK, \bK_{low}, \bK_{diag}}{\text{minimize}} \; J(\bK) + \gamma \| \bK_{low} \|_{*} \nonumber\\
	& \quad\text{subject to} \;  \bK \in \cF, \nonumber\\
	&\quad\quad\quad\quad\quad\; \bK = \bK_{diag} + \bK_{low},
\end{align}
\normalsize
where the nuclear norm is used for the regularization term. By considering the rank of $\bK_{low}$ as a constraint, we can have
\par\noindent\small
\begin{align}\label{eq:LQR_prob3}
	&\underset{\bK, \bK_{low}, \bK_{diag}}{\text{minimize}} \; J(\bK) \nonumber\\
	& \quad\text{subject to} \;  \bK \in \cF, \nonumber\\
	&\quad\quad\quad\quad\quad\; \bK = \bK_{diag} + \bK_{low}, \nonumber \\
	&\quad\quad\quad\quad\quad\;  rank(\bK_{low}) = r
\end{align}
\normalsize
where $rank(\cdot)$ represents the rank of a matrix.

Our goal in this paper is to find a feedback matrix $\bK$ whose decomposition is expressed as a (block) diagonal matrix plus a low-rank matrix by solving the low-rank LQR optimal control design problem \eqref{eq:LQR_prob2} or \eqref{eq:LQR_prob3} . In the next section, we will introduce the standard LQR control design which can provide minimum LQR cost but with heavy communication links, and the sparse LQR control design which can provide a trade-off solution between the LQR cost and the number of communication links in the control of a distributed system.

\section{Previous Research on LQR Control Design}
\label{sec:review}
By setting $\gamma$ to 0 in \eqref{eq:LQR_prob}, the optimization problem \eqref{eq:LQR_prob} becomes the standard LQR optimal control design problem. The standard LQR optimal control design problem has been studied for several decade. Since there is no regularization term for $\bK$, it can provide a feedback matrix $\bK$ with the minimum LQR cost. However, the feedback matrix obtained from this standard LQR control design is normally a dense matrix. In the aspect of communication network, we have normally heavy communication links among agents in the control of a distributed system, which requires lots of communication.

To reduce the number of communcation links in a network, the sparse LQR optimal control design problem or its variation has been studied in previous research such as  \cite{lin2013design,cho2022iterative,lian2017game,cho2020communication} and reference therein. In order to have a sparse feedback matrix which represents the reduction of the number of communication links in a network, for the regularization term $G(\bK)$, element-wise $\ell_1$ norm or its variation were considered, e.g., $G(\bK) = \| \bK \|_1$, or $\ell_1$ norm of off-diagonal matrix of $\bK$, or column-wise $\ell_1$ norm. To solve the sparse LQR optimal control problem, the previous research considered ADMM technique in \cite{lin2013design}, Iterative Shrinkage Thresholding Algorithm (ISTA) in \cite{cho2022iterative}, and Gradient Support Pursuit (GraSP) in \cite{lian2017game}, which can successfully provide a trade-off solution between the LQR cost and the level of sparsity on feedback matrix $\bK$.  

However, we have a question about whether the reduction of the number of communication links can be beneficial to reducing the total energy consumed in a distributed system. To answer this question, in the next sections, let us introduce why low-rank LQR optimal control design can play an important role in the reduction of the total energy consumption against the standard and the sparse LQR control designs, especially in a distributed system over a wireless communication network. 

\section{Low-rank LQR Optimal Control Design}
\label{sec:algo}
In this section, we will introduce the interpretation of the low-rank $\bK_{low}$ in the control of a distributed system. Before the introduction, it is noteworthy that we decompose the feedback matrix $\bK$ into $\bK_{diag}$ and $\bK_{low}$, where $\bK_{diag}$ has only non-zero entries in diagonal (or block-diagonal) which can be linked to a feedback loop in each individual agent, i.e., internal feedback, and $\bK_{low}$ is related to communication links among agents which we would like to reduce. Then, in order to see the physical meaning of the low-rank feedback matrix $\bK_{low}$ in communication, let us consider rank-1 case first, i.e., $rank(\bK_{low}) = 1$. When it is rank-1, we can express $\bK_{low} \in \R^{m \times n}$ as follows:
\par\noindent\small
\begin{align}
\bK_{low} = \begin{bmatrix}a_1 \\ a_2 \\ \vdots \\ a_m \end{bmatrix} \begin{bmatrix}b_1 & b_2 & \cdots & b_n \end{bmatrix},
\end{align}
\normalsize
where $a_i, b_j$, $i=1,...,m$, $j=1,...n$ are arbitrary numbers. Then, the feedback signal $\bu(t)$ is expressed as follows:
\par
\begin{figure}[t]
    \centering
  	\includegraphics[scale=0.55]{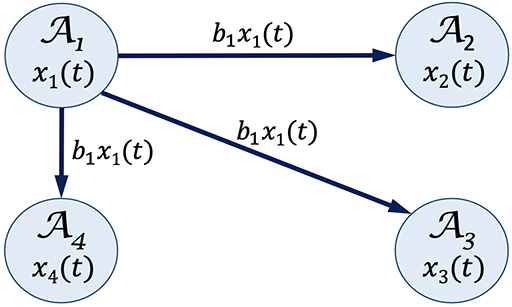}\\
    \caption{Illustration of external feedback, $\bK_{low}$, signals from agent 1 to other agents at time $t$ with a scalar factor $b_1$.}
    \label{fig:feedback_signal_agent1}
\end{figure}

\noindent\small
\begin{align*}
\bu(t) &= -\bK \bx(t) = -(\bK_{diag} + \bK_{low}) \bx(t)\\
		  &= -\bK_{diag}\bx(t) - \begin{bmatrix}a_1 \\ a_2 \\ \vdots \\ a_m \end{bmatrix} \begin{bmatrix}b_1 & b_2 & \cdots & b_n \end{bmatrix} \bx(t)\\
		  & = -\underbrace{\bK_{diag}\bx(t)}_{\text{Internal feedback}} - \underbrace{ \underbrace{ \begin{bmatrix}a_1 \\ a_2 \\ \vdots \\ a_m \end{bmatrix} b_1x_1(t)}_{(A)} - \cdots - \underbrace{\begin{bmatrix}a_1 \\ a_2 \\ \vdots \\ a_m \end{bmatrix} b_n x_n(t)}_{(B)}}_{\text{External feedback signals}},
\end{align*}
\normalsize
where $\bx(t) = [x_1(t), x_2(t), \cdots, x_n(t)]^T$, $(A)$ and $(B)$ represent external feedback signals to every agents from the agent $1$ and the agent $n$ respectively. After solving \eqref{eq:LQR_prob}, we can have $\bK_{low}$. At the initial stage (at time 0), each node can share the scale information, i.e., a column vector in $(A)$ or $(B)$ just one time. Then, each agent can have scale information for other agents, and use them through the infinite time period. Since sharing the scale information for each agent is a one-time operation as a part of initialization, the communication burden for sharing scale information can be limited when it is compared to communication burden during control operation in infinite time domain. In the case of Fig. \ref{fig:feedback_signal_agent1}, where $m=4$, and rank-1 $\bK_{low}$, the number of scale information to share in the initial stage is just three, $a_2$, $a_3$, and $a_4$. 

When we have a distributed control system over a wireless network, agents share communication channels in a shared medium broadcast pattern, which spread every direction over space. Therefore, for the term $(A)$, the agent 1, i.e., $\cA_1$, can share its scaled state $b_1x_1(t)$ at time $t$ in the broadcasting manner, as shown in Fig. \ref{fig:feedback_signal_agent1}. Namely, we do not need to send the state of agent 1, i.e., $x_1(t)$, to each agent one by one, which will cause severe communication delay in control over wireless and may cause performance deterioration and/or possibly the instability of the system due to the delay \cite{liu2009study,xiao2000control,yi2010eigenvalue,Bahavarnia2017Sparse,moelja2005h2}. In terms of consumed energy in communication, since the power density in wireless signal is proportional to the inverse square of the distance \cite{tse2005fundamentals}, the transmission power can be determined by maximum distance. In the case of Fig. \ref{fig:feedback_signal_agent1}, it is the distance between agent 1, $\cA_1$,  and agent 3, $\cA_3$, and the distance can be related to energy consumption in communication. Since in the rank-1 structure, agent 1 can send its state in a broadcast manner to every other agents one time with the power that can reach agent 3, every other node on a wireless communication network can have the state of the agent 1. In contrast, the standard or the sparse LQR control design, it is required to separately send the state of an agent  to other agents. Therefore,  in the standard or the sparse LQR control design, energy consumption and delay in communication can be much larger than those of the low-rank LQR control design on a wireless network. Basically, thanks to the low-rank structure of the feedback matrix, $\bK_{low}$, we can reduce the communication delay as well as communication energy compared to the case separately sharing the state information from one agent to anther. Additionally, in a wireless network, the communication delay caused by sharing channel can be minimized by using orthogonal carrier signals among agents.

In the case of $\bK_{low}$ in rank-$r$, we can express $\bK_{low}$ as
\par\noindent\small
\begin{align}
\bK_{low} = \sum_{k=1}^r \ba_k\bb_k^T,
\end{align}
\normalsize
where $\ba_k \in \R^{m \times 1}$ and $\bb_k \in \R^{n \times 1}$ are arbitrary vectors. For the feedback signal  $\bu(t)$, we have
\par\noindent\small
\begin{align*}
\bu(t) &= -\bK \bx(t) = -(\bK_{diag} + \bK_{low}) \bx(t)\\
		  &= -\bK_{diag}\bx(t) -  \sum_{k=1}^r \ba_k \bb_k^T \bx(t)\\
		  & = -\underbrace{\bK_{diag}\bx(t)}_{\text{Internal feedback}} - \underbrace{ \underbrace{  \sum_{k=1}^r \ba_k b_{k,1} x_1(t)}_{(A)} - \cdots - \underbrace{\sum_{k=1}^r \ba_k b_{k,n} x_n(t)}_{(B)}}_{\text{External feedback signals}},
\end{align*}
\normalsize
where $b_{k,i}$ represents the $i$-th element of the vector $\bb_k$. In this case, even though the size of data, i.e., scale information, at the initial stage, is increased, with low-rank feedback matrix, the data to share at the early stage can also be limited in the similar manner to the rank-1 case, when it is compared to the communication energy in infinite time domain. The number of scale information to share is $O(mr)$, where $m$ is the number of agents.


In the next sub-section, let us introduce the ADMM-based algorithm to solve the low-rank LQR control design problems introduced in \eqref{eq:LQR_prob2} and  \eqref{eq:LQR_prob3} .

\subsection{ADMM-based Algorithm to Solve Low-rank LQR Optimal Control Design Problem}
In order to solve the low-rank LQR optimal control peoblem \eqref{eq:LQR_prob2},  we can use the ADMM technique \cite{boyd2015alternating}. Since the objective function is already separable between $J(\bK)$ and $\|\bK_{low} \|_{*}$, from \eqref{eq:LQR_prob2}, we have the augmented Lagrangian $L_a(\bK, \bK_{diag}, \bK_{low}, \bLambda)$, where $\bLambda$ is a dual variable, as follows:
\par\noindent\small
\begin{align}\label{eq:La}
& L_a(\bK, \bK_{diag}, \bK_{low}, \bLambda) \nonumber\\
& =  J(\bK) + \gamma \| \bK_{low} \|_{*}  +  \langle \bK  -  \bK_{diag}  -   \bK_{low}, \bLambda \rangle \nonumber \\
& \quad\quad + \frac{\rho}{2}\| \bK - \bK_{diag} -  \bK_{low} \|^2_F. 
\end{align}
\normalsize
Then, with the augmented Lagrangian, we can have the following steps for updating variables in ADMM:
\par\noindent\small
\begin{align}
&\bK^{(t+1)} = \underset{\bK \in \cF}{\text{argmin}}\; L_a(\bK, \bK_{diag}^{(t)}, \bK_{low}^{(t)}, \bLambda^{(t)}) \label{eq:ADMM_step_K}\\
&\bK_{diag}^{(t+1)} = \underset{\bK_{diag}}{\text{argmin}}\; L_a(\bK^{(t+1)}, \bK_{diag}, \bK_{low}^{(t)}, \bLambda^{(t)}) \label{eq:ADMM_step_K_diag}\\
&\bK_{low}^{(t+1)} = \underset{\bK_{low}}{\text{argmin}}\; L_a(\bK^{(t+1)}, \bK^{(t+1)}_{diag}, \bK_{low}, \bLambda^{(t)})\label{eq:ADMM_step_K_low}\\
&\bLambda^{(t+1)} = \bLambda^{(t)} + \rho( \bK^{(t+1)} - \bK_{diag}^{(t+1)} - \bK_{low}^{(t+1)}), \label{eq:ADMM_step_Lambda}
\end{align}
\normalsize
where the super-script $(t)$ and $(t+1)$ are used to indicate the $t$-th and $(t+1)$-th iteration number respectively.   We run the aforementioned updating steps until the following stopping criteria meets:
\par\noindent\small 
\begin{align}
&\| \bK^{(t+1)}  - \bK_{diag}^{(t+1)} - \bK_{low}^{(t+1)} \|_F \leq \epsilon_{pri}, \label{eq:stop_cond1}\\
&\| \bK_{diag}^{(t+1)}+  \bK_{low}^{(t+1)}    - \bK_{diag}^{(t)}  - \bK_{low}^{(t)}  \|_F \leq \epsilon_{dual},\label{eq:stop_cond2}\\
& \bK_{diag} + \bK_{low} \in \cF,
\end{align}
\normalsize
where $(\bK^{(t+1)}  - \bK_{diag}^{(t+1)} - \bK_{low}^{(t+1)})$ and $(\bK_{diag}^{(t+1)}+  \bK_{low}^{(t+1)}    - \bK_{diag}^{(t)}  - \bK_{low}^{(t)})$ represent the primal residual and the dual residual at the $(t+1)$ iteration. And correspondingly $\epsilon_{pri}$ and $\epsilon_{dual}$ are small feasibility tolerances for the primal and dual residuals. 

In the detailed step, for \eqref{eq:ADMM_step_K}, we calculate the following optimization problem:
\par\noindent\small
\begin{align}\label{eq:minimize_K}
	&\underset{\bK \in \cF, \bP}{\text{minimize}} \; \Tr(\bB_2^T \bP \bB_2) + \langle \bK , \bLambda^{(t)} \rangle + \frac{\rho}{2} \| \bK - \bK_{diag}^{(t)} - \bK_{low}^{(t)} \|^2_{F} \nonumber\\
	& \text{subject to} \;  (\bA - \bB_1 \bK)^T \bP + \bP(\bA - \bB_1 \bK) + \bQ + \bK^T \bR \bK = 0.
\end{align}
\normalsize
In order to obtain the gradient over $\bK$, by introducing a new variable $\bL$ which needs to satisfy the following Lyapunov equation: 
\begin{align}\label{eq:Lyap_L}
	(\bA - \bB_1 \bK) \bL + \bL( \bA - \bB_1 \bK)^T + \bB_2 \bB_2^T = 0,
\end{align}
we can have the following gradient of the objective function as follows \cite{rautert1997computational}:
\par\noindent\small
\begin{align}
&\grad_{\bK} L_a(\bK, \bK^{(t)}_{diag}, \bK_{low}^{(t)}, \bLambda^{(t)})  \\
& = 2[ \bB_1^T \bP(\bK) - \bR \bK] \bL(\bK) + \bLambda^{(t)} + \rho ( \bK - \bK_{diag}^{(t)} - \bK_{low}^{(t)}),\nonumber
\end{align}
\normalsize
where $\bP(\bK)$ and $\bL(\bK)$ are functions of $\bK$ satisfying the Lyapunov equations \eqref{eq:Lyap_P} and \eqref{eq:Lyap_L} respectively. By considering the first-order condition at an optimal solution, i.e., gradient at an optimal solution needs to be zero, we can obtain an optimal solution for $\bK^{(t+1)}$ by using a well-known fixed-point iterative method, so-called Anderson-Moore algorithm \cite{rautert1997computational}. 

For $\bK^{(t+1)}_{diag}$, by taking into account the first-order condition at $\bK^{(t+1)}_{diag}$, we can find an optimal solution for $\bK^{(t+1)}$ satisfying the following the first-order condition:
\par\noindent\small
\begin{align*}
&\bLambda^{(t)} + \rho (\bK^{(t+1)} - \bK_{diag}^{(t+1)} - \bK_{low}^{(t)} ) = \bm{0}.
\end{align*}
\normalsize
And by considering the diagonal (or block diagonal) structure of $\bK_{diag}$, we have 
\par\noindent\small
\begin{align}\label{eq:ADMM_update_K_diag_detail}
\bK^{(t+1)}_{diag} = \text{diag}_{trim}\bigg( \frac{1}{\rho}\bLambda^{(t)} + \bK^{(t+1)} - \bK_{low}^{(t)} \bigg),
\end{align}
\normalsize
where $\diag_{trim}(\cdot)$ is an operator to make a diagonal matrix by taking the elements in diagonal only and setting the off-diagonal to zero. 

Then, for $\bK_{low}^{(t+1)}$, we can solve the following optimization problem:
\par\noindent\small
\begin{align}\label{eq:minimize_K2}
&\bK_{low}^{(t+1)} \\
	&=\underset{\bK_{low}}{\text{argmin}} \; \gamma\| \bK_{low} \|_{*} +  \frac{\rho}{2} \| \bK^{(t+1)} - \bK_{diag}^{(t+1)} + \frac{1}{\rho} \bLambda^{(t)} -  \bK_{low} \|^2_{F}. \nonumber
\end{align}
\normalsize
The level of the rank in the optimal solution $\bK_{low}^{(t+1)}$ will be determined by the ratio parameter $\rho/\gamma$. If $\rho/\gamma$ is large, then, in order to reduce the misfit error in the Frobenius norm, the optimal solution will be clear to $(\bK^{(t+1)} - \bK_{diag}^{(t+1)} + \frac{1}{\rho} \bLambda^{(t)})$, and possibly not have low-rank. If  $\rho/\gamma$ is small enough, then, by allowing more misfit error in the Frobenius norm, we can have a low-rank solution. Basically, the ratio parameter $\rho/\gamma$ determines the level of the rank in the optimal solution $\bK_{low}^{(t+1)}$. And the low-rank solution, $\bK_{low}^{(t+1)}$, is expressed as follows \cite{Hu2013Fast}:
\par\noindent\small
\begin{align} \label{eq:ADMM_update_K_low_detail}
\bK_{low}^{(t+1)} = \bU \cS_{\rho/\gamma}(\bSigma)\bV^T,
\end{align} 
\normalsize
where $\bU \bSigma \bV^T$ is the Singular Value Decomposition (SVD) of $(\bK^{(t+1)} - \bK_{diag}^{(t+1)} + \frac{1}{\rho} \bLambda^{(t)})$, and $\cS_{\rho/\gamma}(\bSigma)$ is the shrinkage-threshold operator $\R^{m \times n} \rightarrow \R^{m \times n}$ stated as follows:
\par\noindent\small
\begin{align}\label{eq:soft_thresholding}
	\cS_{\rho/\gamma}(\bSigma)_{i,j} = \max \{\Sigma_{i,j} - \rho/\gamma, 0 \}.
\end{align}  
\normalsize
Notice that $\bSigma$ is a diagonal matrix having singular values in diagonal. We call \eqref{eq:soft_thresholding} as soft-thresholding operation. 

Additionally, in order to have $\bK_{low}^{(t+1)}$ in rank-$r$, we can calculate the SVD of $(\bK^{(t+1)} - \bK_{diag}^{(t+1)} + \frac{1}{\rho} \bLambda^{(t)})$; and by taking the first $r$ largest singular values and corresponding singular vectors, we can obtain a $\bK_{low}^{(t+1)}$ matrix in rank-$r$, which is expressed as follows:
\par\noindent\small
\begin{align}\label{eq:hard_thresholding}
	\cH_{r}(\bSigma) = \diag([\sigma_1, \sigma_2, ..., \sigma_r]),
\end{align}  
where $\sigma_i$ is the $i$-th singular value of  $(\bK^{(t+1)} - \bK_{diag}^{(t+1)} + \frac{1}{\rho} \bLambda^{(t)})$ in descending order, and $\diag(\cdot)$ represents the diagonal function making a diagonal matrix for a given vector by placing the vector elements in diagonal. We call \eqref{eq:hard_thresholding} as hard thresholding operation. And then,  $\bK_{low}^{(t+1)}$ matrix in rank-$r$ is obtained as follows:
\par\noindent\small
\begin{align} \label{eq:ADMM_update_K_low_detail_hard}
\bK_{low}^{(t+1)} = \bU_{1:r} \cH_{r}(\bSigma)\bV_{1:r}^T,
\end{align} 
\normalsize
where $\bU_{1:r}$ and $\bV_{1:r}$ are the partial matrices of $\bU$ and $\bV$ respectively by taking columns from the $1$-st column to the $r$-th column. \eqref{eq:ADMM_update_K_low_detail} and \eqref{eq:ADMM_update_K_low_detail_hard} can be used to solve the low-rank LQR control design problems \eqref{eq:LQR_prob2} and \eqref{eq:LQR_prob3} respectively.  We summarize the updating steps in Algorithm \ref{alg:ADMM}. We will then introduce the brief analysis of these algorithms on optimality in the next subsection.

\begin{algorithm}[t]
  \caption{Alternating Direction Method of Multipliers (ADMM) for Low-rank LQR Optimal Control Design}
  \label{alg:ADMM}
  \SetAlgoLined
  \SetKwRepeat{Do}{do}{while}%
{\small
   \KwIn{ $\bA$, $\bB_1$, $\bQ \succeq 0$, $\bR \succ 0$, $\rho$, $\gamma$ (or rank $r$), $\epsilon_{pri}$, $\epsilon_{dual}$, $MaxItr$ }
   \textbf{Initialization}: $\bK^{(0)} \leftarrow$ a stable dense matrix $\bK \in \cF$ from a solution to LQR problem with $\gamma = 0$   \par
   \For { $t=0$  \KwTo $MaxItr$ }
   {
   		   $\bK^{(t+1)}$ $\leftarrow$ solution of \eqref{eq:ADMM_step_K} \par
   		   $\bK_{diag}^{(t+1)}$ $\leftarrow$ solution of \eqref{eq:ADMM_update_K_diag_detail} \par
    	   $\bK_{low}^{(t+1)}$ $\leftarrow$ solution of \eqref{eq:ADMM_update_K_low_detail} for soft-thresholding (or \eqref{eq:ADMM_update_K_low_detail_hard} for hard-thresholding)\par
    	   $\bLambda^{(t+1)}$ $\leftarrow$ solution of \eqref{eq:ADMM_step_Lambda} \par    		   
			\If { \eqref{eq:stop_cond1},  \eqref{eq:stop_cond2}, and $(\bK^{(t+1)}_{diag} + \bK^{(t+1)}_{low}) \in \cF$ hold true } 
			{
				break
			} 			
}
  \KwOut{ Structured feedback matrix $\bK$ }
  }
\end{algorithm}


\subsection{Optimality of ADMM-based Algorithm to Solve the Low-rank LQR Optimal Control Design Problem}
\label{sec:analysis}

Let us introduce the analysis of the ADMM-based algorithm to solve the low-rank LQR optimal control design problem. Especially, for the optimality of the ADMM-based algorithm, an optimal solution needs to satisfy the following primal feasibility condition:
\par\noindent\small
\begin{align}
\bK^{\star} - \bK^{\star}_{diag} - \bK_{low}^{\star} = \bm{0}, \label{eq:primal_feasible}
\end{align}
\normalsize
where the super-script $\star$ represents the optimal solution to the problem in \eqref{eq:LQR_prob2}. With the given augmented Lagrangian \eqref{eq:La}, for dual feasible conditions over $\bK^{\star}$ and $\bK_{low}^{\star}$, the gradient value of the objective function of \eqref{eq:LQR_prob2} at the optimal point needs to be zero. Thus, we have 
\par\noindent\small
\begin{align}
	&\bm{0} \in \partial J(\bK^{\star}) + \bLambda^{\star},\label{eq:FO_K_star}\\
	&\bm{0} \in \partial \| \bK_{low}^{\star} \|_{*} - \bLambda^{\star},\label{eq:FO_F_star}
\end{align}
\normalsize
where $\partial$ represents the subdifferential operator \cite{boyd2015alternating}. 


In the ADMM step for $\bK_{low}^{(t+1)}$, $\bK_{low}^{(t+1)}$ minimizes $L_a(\bK^{(t+1)}, \bK^{(t+1)}_{diag}, \bK_{low}, \bLambda^{(t)})$. Hence, we have the following condition for $\bK_{low}^{(t+1)}$:
\par\noindent\small
\begin{align*}
\bm{0} & \in \partial \| \bK_{low}^{(t+1)} \|_{*} - \bLambda^{(t)} - \rho ( \bK^{(t+1)} - \bK^{(t+1)}_{diag} - \bK_{low}^{(t+1)} ) \\
		    & = \partial \| \bK_{low}^{(t+1)} \|_{*} - \bLambda^{(t+1)},
\end{align*}
\normalsize
where the equality is obtained from the ADMM step in \eqref{eq:ADMM_step_Lambda}. This indicates that with $\bK^{(t+1)}$ and $\bLambda^{(t+1)}$, the dual feasible condition \eqref{eq:FO_F_star} always holds. 

Then, in order to check the optimality conditions of \eqref{eq:primal_feasible} and \eqref{eq:FO_K_star} for $\bK^{(t+1)}$, which minimizes $L_a(\bK, \bK^{(t)}_{diag}, \bK_{low}^{(t)}, \bLambda^{(t)})$, we have
\par\noindent\small
\begin{align*}
\bm{0} & \in \partial J(\bK^{(t+1)}) + \bLambda^{(t)} + \rho ( \bK^{(t+1)} - \bK^{(t)}_{diag} - \bK_{low}^{(t)} ) \\
			&  =  \partial J(\bK^{(t+1)}) + \bLambda^{(t)} + \rho ( \bK^{(t+1)} - \bK^{(t)}_{diag} - \bK_{low}^{(t)} ) \\
			& \quad\quad\quad \quad + \rho (\bK^{(t+1)}_{diag} + \bK_{low}^{(t+1)}) - \rho (\bK^{(t+1)}_{diag} + \bK_{low}^{(t+1)})\\
 		    & = \partial J(\bK^{(t+1)}) + \bLambda^{(t+1)} \\
 		    & \quad\quad\quad \quad + \rho( \bK^{(t+1)}_{diag} + \bK_{low}^{(t+1)} - \bK_{diag}^{(t)} - \bK_{low}^{(t)}),
\end{align*}
\normalsize
where the term $\rho( \bK^{(t+1)}_{diag} +\bK_{low}^{(t+1)} - \bK_{diag}^{(t)} - \bK_{low}^{(t)})$ can be thought of as a residual that needs to go to zero as the iteration step goes. Additionally, for the primal feasible condition \eqref{eq:primal_feasible}, we check the primal residual which is defined as $\bK^{(t+1)}  - \bK_{diag}^{(t+1)} - \bK_{low}^{(t+1)}$ in our stopping criteria. The residuals converge to zero through the ADMM updating steps, i.e., minimizing the augmented Lagrangian and updating the dual variable. Therefore, with small feasibility tolerances, $\epsilon_{pri}$ and $\epsilon_{dual}$, the estimated solution obtained from the ADMM updating step can be considered to be close to an optimal solution. Additionally, the low-rank solution, i.e., $\bK_{diag} + \bK_{low}$, needs to be in the feasible set $\cF$ as another primal feasibility condition. Hence, we check the condition as a part of our stopping criteria. 


\section{Numerical experiments}
\label{sec:experiment}
In the numerical experiments, we simulate various distributed multi-agent control system models where each agent has $(x,y)$ coordinates as its location on a plane. By considering the characteristics of wireless communication, namely, the communication power density is inversely proportional to the square of the distance, and wireless signal can spread every direction, we run numerical experiments to compare the low-rank LQR optimal control design against the sparse and the standard LQR optimal control designs on a distributed multi-agent control system model. 

For the distributed multi-agent control system model, we deal with the following second order system model having coupling with other agents through the exponentially decaying function of the Euclidean distance between any two nodes:
\par\noindent\small
\begin{align} \label{eq:signal_model}
	\begin{bmatrix} \dot{x}_{i}(t)_1 \\ \dot{x}_i(t)_2 \end{bmatrix} = & \begin{bmatrix} 1 & 1 \\ 1  & 3 \end{bmatrix} \begin{bmatrix} x_{i}(t)_{1} \\ x_i(t)_{2} \end{bmatrix} \nonumber\\ &+ \sum_{i\neq j} e^{d(i,j)} \begin{bmatrix} x_{j}(t)_{1} \\ x_j(t)_{2} \end{bmatrix} + \begin{bmatrix} 0\\ 1 \end{bmatrix} \bigg(w_i(t) + u_i(t)\bigg),
\end{align}
\normalsize
where the subscript $i$ represents the $i$-th agent having two states $x_i(t)_1$ and $x_i(t)_2$, where $i=1,2,...,n$, $w_i(t)$ and $u_i(t)$ are the disturbance and input signals of the $i$-th agent respectively, and $d(i,j)$ represents the Euclidean distance between the $i$-th agent and $j$-th agent. We vary the number of agents $N$ from 10 to 20 and choose the locations of agents on a $10 \times 10$ plane uniformly at random. For both $\bQ$, and $\bR$, we use identity matrices.

For the initial point of the algorithms for both low-rank LQR design and sparse LQR design, we use the Linear-Quadratic Regulator (LQR) Matlab function to have the standard LQR control design solution, which normally provides a dense feedback matrix $\bK$, but with minimum LQR cost. We denote it as $J_{\text{stand}}$.

\subsection{Communication scenario 1: Fixed communication power} 
\begin{figure}[t]
    \centering
    \includegraphics[scale=0.9]{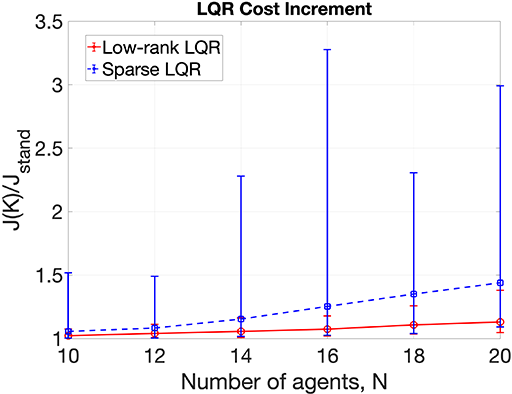}\;\;
    \caption{Comparison in LQR cost increment between the low-rank LQR design and the sparse LQR design in terms of the standard LQR cost, denoted by $J_{\text{stand}}$, under fixed communication power scenario. }
    \label{fig:comp_scenario1}
\end{figure}

In this scenario, we consider a case where each agent transmits or broadcasts its states at each time with fixed transmission power. We assume that the all agents are reachable each other on a wireless network with the transmission power. Since communication power of an agent at each time is the same for all agents, for communication burden, we take into account total communication energy as power $\times$ time, which can be estimated by the number of communication attempt. For the controller based on the low-rank LQR design, we choose $\bK_{low}$ to be rank-1. With the feedback controller $\bK = \bK_{diag} + \bK_{low}  \in \R^{m \times n}$, if there is no communication error and no zero column vector in $\bK_{low}$, it requires $n$ communication attempts. To have the same number of minimum communication attempts in the feedback matrix based on sparse LQR design for comparison purpose, we adjust the parameter $\gamma$ in \eqref{eq:LQR_prob} to have $n$ communication links. Thus, in the case of no error in communication, we can expect to have same communication burden between the low-rank LQR design and the sparse LQR-design. Hence, we compare the LQR cost increment of the two different designs from the standard LQR cost, i.e., $J_{\text{stand}}$, by varying the number of agents in the system. We run this simulation for hundred trials with randomly chosen nodes. The number of nodes, i.e., agents, are varied from 10 to 20. Fig. \ref{fig:comp_scenario1} shows the LQR cost increment from the standard LQR design by calculating $J(\bK)/J_{\text{stand}}$, where $\bK$ is determined by the feedback controller design. Red solid and blue dotted lines represent the low-rank LQR design and the sparse LQR design respectively. A vertical line represents minimum and maximum LQR cost increment among 100 random trials. As shown in Fig. \ref{fig:comp_scenario1}, under the same communication burden, the increment rate of the LQR cost in the low-rank feedback controller design is significantly smaller than the increment rate in the sparse feedback controller design. It is noteworthy that the more LQR cost is incremented, the more energy consumption in control is expected.

Unfortunately, we normally have error in communication. When we denote the probability of error in communication as $P_e$, the total error probability in the rank-1 feedback controller design is $1-(1-P_e)^{mn}$, while the total error probability in the sparse feedback controller design having $n$ communication links is $1-(1-P_e)^{n}$. Because of the parameter $m$, the total error probability in communication on the low-rank feedback controller can be larger than that of the sparse feedback controller. This is because in the low-rank LQR design having rank-1, we have $m$ times larger number of communication links than that of the sparse LQR design. And the simulation shown in Fig. \ref{fig:comp_scenario1} is a  special case, when $P_e = 0$. That is one drawback of the low-rank LQR design. In order to reflect the cases of communication error and noise, we introduce the next simulation scenarios.

\subsection{Communication scenario 2: System endurance against communication noise}
\begin{figure*}[t]
    \centering
    \subfloat[Standard LQR design]{\includegraphics[scale=0.63]{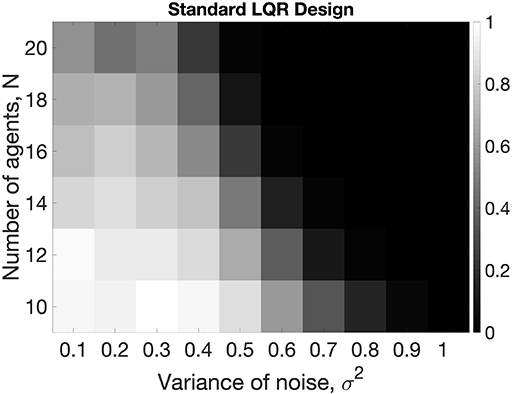}}\;\;
    \subfloat[Sparse LQR design]{\includegraphics[scale=0.63]{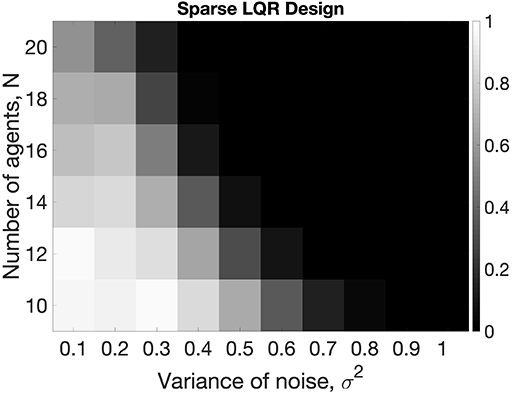}}\;\;
    \subfloat[Low-rank LQR design]{\includegraphics[scale=0.63]{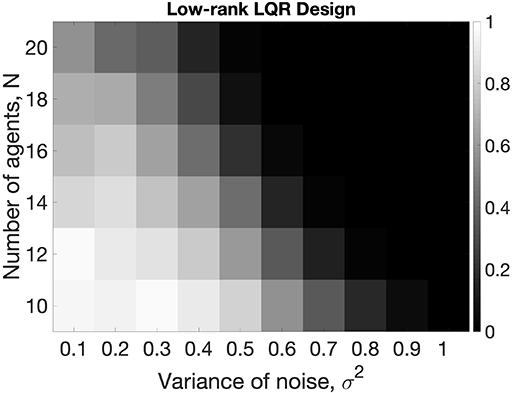}}
    \caption{Probability of success that corrupted feedback matrix  by noise is in $\cF$. Comparison among (a) Standard LQR design, (b) Sparse LQR design, and (c) Low-rank LQR design in term of system stability endurance against noise in communication.  }
    \label{fig:comp_scenario2}
\end{figure*}

In this scenario, we check the system endurance against communication noise, since communication noise is inevitable. We take into account communication noise following i.i.d Gaussian distribution $\cN(0,\sigma^2)$ in each commutation link. With the two feedback matrices obtained from the low-rank LQR design and the sparse LQR design, at each time of transmission of states from one agent to another, we add the Gaussian noise to transmitted signals. For that, we generate Gaussian noise for each element in off-diagonal of a feedback matrix, and add the noise to the feedback matrix. And then, we check whether the feedback matrix is in $\cF$ or not. In the next round of random test, we generate another noise for each element in off-diagonal of the feedback matrix, and check the stability of the feedback matrix.  For a feedback matrix, we run hundred trials with randomly chosen noise and count the occurrences when the corrupted feedback matrix by noise is in $\cF$. If the corrupted feedback matrix is in $\cF$, then, we consider it as success. We run the trials for 100 different feedback matrices for the low-rank LQR design and the sparse LQR design respectively. Therefore, for each parameter setup shown in Fig. \ref{fig:comp_scenario2}, we calculate the probability of success in $100 \times 100$ random cases. Black and white boxes represent probability 0 and 1 respectively.  The variance of noise $\sigma^2$ is varied from $0.1$ to $0.9$. As shown in Fig. \ref{fig:comp_scenario2}, the low-rank LQR design has similar robustness against noise to the standard LQR design, and has more robustness than the sparse LQR design.

\subsection{Communication scenario 3: Security under cyber attack}
Security is another critical issue to be considered in the control of distributed systems. By taking into account a scenario of cyber attack, in this simulation, we forcedly remove some communication links and check the system stability between the low-rank LQR control design and the sparse LQR control design. Hence, through this simulation, we compare the performance of the low-rank LQR control design against that of the sparse LQR control design in terms of system stability endurance against cyber attack. More specifically, under the assumption that any data or signal cannot go through communication links under attack, we measure the margin of the feedback control system in stability by varying the number of communication links under attack from 1 to 10. The communication links under attack are randomly chosen among the required communication links. Hence, through this scenario, we investigate the system robustness against cyber attack. In the aspect of noise, this scenario can be thought of as hard noise case, i.e., completely fail in communication for some links, while the Communication scenario 2 deals with a soft noise case.  We run simulations and check the performance of the system  robustness against cyber attack between the low-rank LQR control design and the sparse LQR control design. Depending on the number of removed communication links in maximum allowing the system to be stable, the performance in system robustness against cyber attack is evaluated.

\begin{figure*}[t]
    \centering
    \subfloat[Standard LQR design]{\includegraphics[scale=0.63]{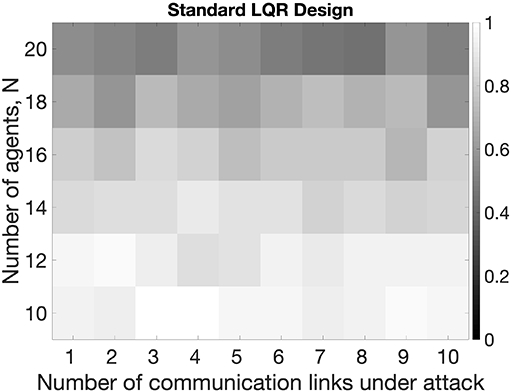}}\;\;
    \subfloat[Sparse LQR design]{\includegraphics[scale=0.63]{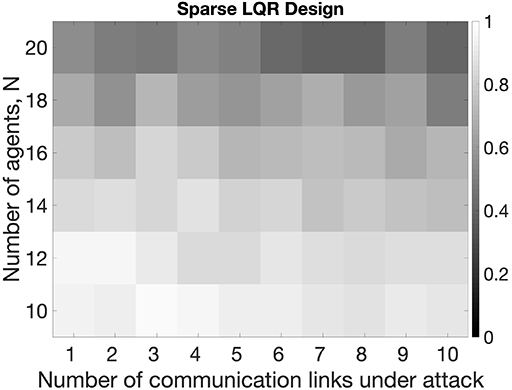}}\;\;
    \subfloat[Low-rank LQR design]{\includegraphics[scale=0.63]{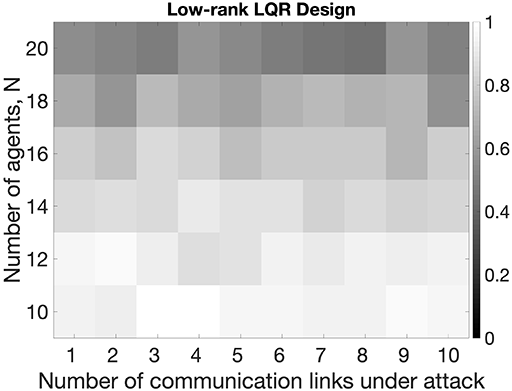}}
    \caption{Comparison between the low-rank LQR design and the sparse LQR design in term of system stability endurance against cyber attack.  Comparison among (a) Standard LQR design, (b) Sparse LQR design, and (c) Low-rank LQR design in term of system stability endurance against cyber attack. }
    \label{fig:comp_scenario3}
\end{figure*}

With the two feedback matrices obtained from the low-rank LQR design and the sparse LQR design, at each time of transmission of states from one agent to another, we assume that we have communication links under attack. The number of links under attack, $l$, is varied from 1 to 10. We choose $l$ elements in off-diagonal of a feedback matrix uniformly at random for the links having cyber attack, and make the chosen elements to zero. And then we check whether the feedback matrix is in $\cF$ or not. In the next round of random test, we choose again $l$ elements uniformly at random, and check the stability of the feedback matrix after setting the chosen elements to zero.  For a feedback matrix, we run hundred random trials on cyber attack and count the occurrences where the modified feedback matrix is in $\cF$. If the modified feedback matrix is in $\cF$, then, we consider it as success. For hundred different feedback matrices, we repeat the same scenario. Therefore, for each parameter setup shown in Fig. \ref{fig:comp_scenario3}, we run $100 \times 100$ trials and check the probability of success, where black and white boxes represent the probability of success 0 and 1 respectively. As shown in Fig. \ref{fig:comp_scenario3}, we demonstrate that the low-rank LQR control design can have more stability margin against cyber attack than the sparse LQR control design, and can have similar stability margin to the standard LQR design against cyber attack with much more reduced communication energy.

\subsection{Communication scenario 4: Limited communication energy}
\begin{figure}[t]
    \centering
    \subfloat[Number of links in a critical node in the sparse LQR design]{\includegraphics[scale=0.8]{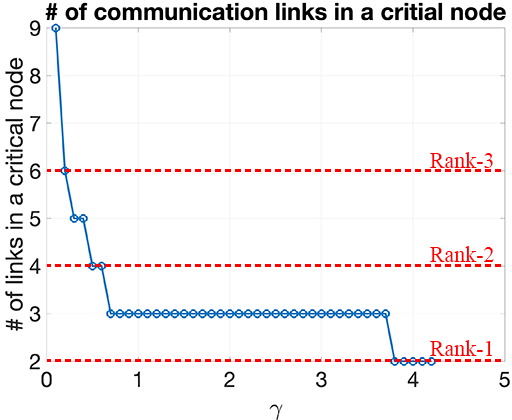}}\\
    \subfloat[LQR cost comparison]{\includegraphics[scale=0.8]{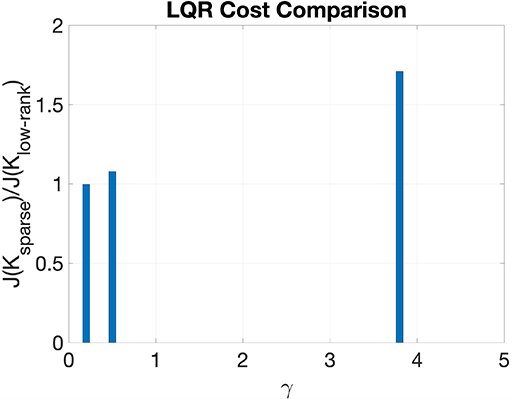}}
    \caption{Comparison between the low-rank LQR design and the sparse LQR design in term of LQR cost by matching the number of communication links in a critical node.  }
    \label{fig:comp_scenario4}
\end{figure}

This scenario considers applications such as a distributed multi-drone, Unmanned Aerial Vehicle (UAV), based system or a distributed multi-robot system, where we have limited communication energy because of agents being battery-powered; namely, each agent has limited energy. We further assume that all agents are reachable each other, and each communication attempt spends the same amount of power. Under this assumption, we compare the ratio of the LQR cost between the sparse LQR design and the low-rank LQR design by matching the number of communication links in a critical node. Namely, in the sparse LQR control design, a node having largest communication links among all nodes is chosen as a critical node. This is because the node needs to send its states to other nodes through communication links. Therefore, in a scenario of UAV, the operation time of the node, i.e., agent, will be the shortest. In contrast, in the low-rank LQR design, every node will have balanced energy consumption in communication. To compare the two different designs, we match the number of communication links in a critical node. Since each agent needs to transmit two states $x(t)_1$ and $x(t)_2$ in \eqref{eq:signal_model}, for $\bK_{low}$ in rank-$r$, we need $2\times r$ numbers of communication transmissions, which is shown in dotted lines in Fig. \ref{fig:comp_scenario4}(a). By considering this scenario, we evaluate the low-rank LQR optimal control design against the sparse LQR control design in terms of LQR cost. As shown in \ref{fig:comp_scenario4}(b), for rank-1 $\bK_{low}$ design, the LQR cost of the sparse LQR design is increased by 71\%, when it is compared to the LQR cost of the low-rank LQR design.

Intuitively, a feedback controller based on the sparse LQR control design can have a critical node, i.e., agent. Hence, we can anticipate that the energy that the critical node has can be consumed much faster than the other agents, since that critical node needs to continuously send its states to other nodes, which can easily jeopardize the whole distributed control system. In contrast, in the low-rank LQR control design, wireless signal is transmitted in a broadcast way. Therefore, even though an agent has lots of communication links to other agents, the number of communication transmissions at the agent is not proportional to the number of communication links, but  it is more related to the communication error. However, in sparse LQR controller, the communication transmissions is linearly proportional to the number of communication links, because the agent needs to separately transmit its states to the connected agents.



\section{Conclusion and discussion}
\label{sec:conclusion}
In this paper, we consider a LQR optimal control design problem with a constraint on communication to optimally controlling distributed systems having multi-agents, which can find applications in smart-grid and multi-agent robot systems. Especially, by considering wireless communication networks and taking advantage of its characteristics, i.e., electromagnetic signal can be spread all directions in a broadcast way, we propose low-rank LQR optimal control design problem and the ADMM-based algorithm to solve the problem. The low-rank LQR control design provides a trade-off solution between the LQR cost and the communication energy in a distributed control system having feedback loops. Through various numerical experiments in different communication scenarios, we demonstrate that the low-rank LQR control design can provide better feedback controller design than that of the sparse LQR optimal control design in terms of energy consumption, system stability margin against noise and error in communication. 

We introduce possible future research directions for low-rank LQR optimal control design as follows:
\begin{itemize}
    \item Designing fast algorithms to solve the proposed low-rank LQR control design problems can be a possible future research topic. 
    \item Finding a way to express the proposed low-rank LQR control design problem into a convex optimization problem such as semi-definite Programming (SDP) formulation is interesting like we have \cite{Balakrishnan1995Connections,Geromel1991On,Ramos2002An,Geromel1996Convex,fazelnia2016convex} for the standard LQR optimal control design problem. By having the convex optimization problem, we can use off-the-shelf solvers, e.g., CVX \cite{cvx}, and have a global solution guaranteed.  
    \item For large-scale distributed control systems, running algorithms to solve the low-rank LQR optimal control problems in limited time can be computationally challenging. Therefore, data-driven approach to design a feedback controller can also be an interesting topic. 
\end{itemize}


\begin{thebibliography}{10}

\bibitem{rautert1997computational}
T.~Rautert and E.~W. Sachs,
\newblock ``Computational design of optimal output feedback controllers,''
\newblock {\em SIAM Journal on Optimization}, vol. 7, no. 3, pp. 837--852,
  1997.

\bibitem{zhou1996robust}
K.~Zhou, J.~C. Doyle, and K.~Glover,
\newblock {\em Robust and optimal control},
\newblock Prentice Hall, Inc., 1996.

\bibitem{peres1994alternate}
P.~L.~D. Peres and J.~C. Geromel,
\newblock ``An alternate numerical solution to the linear quadratic problem,''
\newblock {\em IEEE Transactions on Automatic Control}, vol. 39, no. 1, pp.
  198--202, 1994.

\bibitem{fazelnia2016convex}
G.~Fazelnia, R.~Madani, A.~Kalbat, and J.~Lavaei,
\newblock ``Convex relaxation for optimal distributed control problems,''
\newblock {\em IEEE Transactions on Automatic Control}, vol. 62, no. 1, pp.
  206--221, 2017.

\bibitem{bamieh2002distributed}
B.~Bamieh, F.~Paganini, and M.~A. Dahleh,
\newblock ``Distributed control of spatially invariant systems,''
\newblock {\em IEEE Transactions on automatic control}, vol. 47, no. 7, pp.
  1091--1107, 2002.

\bibitem{de2002convex}
G.~A. De~Castro and F.~Paganini,
\newblock ``Convex synthesis of localized controllers for spatially invariant
  systems,''
\newblock {\em Automatica}, vol. 38, no. 3, pp. 445--456, 2002.

\bibitem{d2003distributed}
R.~D'Andrea and G.~E. Dullerud,
\newblock ``Distributed control design for spatially interconnected systems,''
\newblock {\em IEEE Transactions on automatic control}, vol. 48, no. 9, pp.
  1478--1495, 2003.

\bibitem{bamieh2005convex}
B.~Bamieh and P.~G. Voulgaris,
\newblock ``A convex characterization of distributed control problems in
  spatially invariant systems with communication constraints,''
\newblock {\em Systems \& control letters}, vol. 54, no. 6, pp. 575--583, 2005.

\bibitem{motee2008optimal}
N.~Motee and A.~Jadbabaie,
\newblock ``Optimal control of spatially distributed systems,''
\newblock {\em IEEE Transactions on Automatic Control}, vol. 53, no. 7, pp.
  1616--1629, 2008.

\bibitem{fardad2009optimal}
M.~Fardad, F.~Lin, and M.~R. Jovanovi{\'c},
\newblock ``On the optimal design of structured feedback gains for
  interconnected systems,''
\newblock in {\em Proceedings of IEEE Conference on Decision and Control (CDC)
  held jointly with 2009 28th Chinese Control Conference}, 2009, pp. 978--983.

\bibitem{fardad2014design}
M.~Fardad and M.~R. Jovanovi{\'c},
\newblock ``On the design of optimal structured and sparse feedback gains via
  sequential convex programming,''
\newblock in {\em Proceedings of American Control Conference}. IEEE, 2014, pp.
  2426--2431.

\bibitem{lin2011augmented}
F.~Lin, M.~Fardad, and M.~R. Jovanovic,
\newblock ``Augmented lagrangian approach to design of structured optimal state
  feedback gains,''
\newblock {\em IEEE Transactions on Automatic Control}, vol. 56, no. 12, pp.
  2923--2929, 2011.

\bibitem{fardad2011design}
M.~Fardad and M.~R. Jovanovi{\'c},
\newblock ``Design of optimal controllers for spatially invariant systems with
  finite communication speed,''
\newblock {\em Automatica}, vol. 47, no. 5, pp. 880--889, 2011.

\bibitem{lin2013design}
F.~Lin, M.~Fardad, and M.~R. Jovanovi{\'c},
\newblock ``Design of optimal sparse feedback gains via the alternating
  direction method of multipliers,''
\newblock {\em IEEE Transactions on Automatic Control}, vol. 58, no. 9, pp.
  2426--2431, 2013.

\bibitem{lian2017game}
F.~Lian, A.~Chakrabortty, and A.~Duel-Hallen,
\newblock ``Game-theoretic multi-agent control and network cost allocation
  under communication constraints,''
\newblock {\em IEEE journal on selected areas in communications}, vol. 35, no.
  2, pp. 330--340, 2017.

\bibitem{cho2022iterative}
M.~Cho and A.~Chakrabortty,
\newblock ``Iterative shrinkage-thresholding algorithm and model-based neural
  network for sparse {LQR} control design,''
\newblock in {\em Proceedings of European Control Conference (ECC)}, 2022, pp.
  2311--2316.

\bibitem{cho2020communication}
M.~Cho and A.~S. Abdallah,
\newblock ``Communication-efficient optimal control design for distributed
  control systems in cooperative vehicular networks,''
\newblock in {\em Proceedings of IEEE Vehicular Technology Conference
  (VTC2020-Spring)}, 2020, pp. 1--5.

\bibitem{chen1998linear}
C.-T. Chen,
\newblock {\em Linear system theory and design},
\newblock Oxford University Press, Inc., 1998.

\bibitem{liu2009study}
J.~Liu, A.~Gusrialdi, D.~Obradovic, and S.~Hirche,
\newblock ``Study on the effect of time delay on the performance of distributed
  power grids with networked cooperative control,''
\newblock {\em IFAC Proceedings Volumes}, vol. 42, no. 20, pp. 168--173, 2009.

\bibitem{xiao2000control}
L.~Xiao, A.~Hassibi, and J.~P. How,
\newblock ``Control with random communication delays via a discrete-time jump
  system approach,''
\newblock in {\em Proceedings of the American Control Conference (ACC)}. IEEE,
  2000, vol.~3, pp. 2199--2204.

\bibitem{yi2010eigenvalue}
S.~Yi, P.~Nelson, and A.~Ulsoy,
\newblock ``Eigenvalue assignment via the lambert {W} function for control of
  time-delay systems,''
\newblock {\em Journal of Vibration and Control}, vol. 16, no. 7-8, pp.
  961--982, 2010.

\bibitem{Bahavarnia2017Sparse}
M.~Bahavarnia and N.~Motee,
\newblock ``Sparse memoryless {LQR} design for uncertain linear time-delay
  systems,''
\newblock {\em IFAC-PapersOnLine}, vol. 50, no. 1, pp. 10395--10400, 2017.

\bibitem{moelja2005h2}
A.~A. Moelja and G.~Meinsma,
\newblock ``{H2}-optimal control of systems with multiple i/o delays: Time
  domain approach,''
\newblock {\em Automatica}, vol. 41, no. 7, pp. 1229--1238, 2005.

\bibitem{tse2005fundamentals}
D.~Tse and P.~Viswanath,
\newblock {\em Fundamentals of wireless communication},
\newblock Cambridge university press, 2005.

\bibitem{boyd2015alternating}
N.~Boyd, G.~Schiebinger, and B.~Recht,
\newblock ``The alternating descent conditional gradient method for sparse
  inverse problems,''
\newblock {\em SIAM Journal on Optimization}, vol. 27, no. 2, pp. 616--639,
  2017.

\bibitem{Hu2013Fast}
Y.~Hu, D.~Zhang, J.~Ye, X.~Li, and X.~He,
\newblock ``Fast and accurate matrix completion via truncated nuclear norm
  regularization,''
\newblock {\em IEEE Transactions on Pattern Analysis and Machine Intelligence},
  vol. 35, no. 9, pp. 2117--2130, 2013.

\bibitem{Balakrishnan1995Connections}
V.~Balakrishnan and L.~Vandenberghe,
\newblock ``Connections between duality in control theory and convex
  optimization,''
\newblock in {\em Proceedings of American Control Conference (ACC)}, 1995,
  vol.~6, pp. 4030--4034.

\bibitem{Geromel1991On}
J.~C. Geromel, P.~L.~D. Peres, and J.~Bernussou,
\newblock ``On a convex parameter space method for linear control design of
  uncertain systems,''
\newblock {\em SIAM Journal on Control and Optimization}, vol. 29, no. 2, pp.
  381--402, 1991.

\bibitem{Ramos2002An}
D.~C.~W. Ramos and P.~L.~D. Peres,
\newblock ``An {LMI} condition for the robust stability of uncertain
  continuous-time linear systems,''
\newblock {\em IEEE Transactions on Automatic Control}, vol. 47, no. 4, pp.
  675--678, 2002.

\bibitem{Geromel1996Convex}
J.~C. Geromel, P.~L.~D. Peres, and S.~R. Souza,
\newblock ``Convex analysis of output feedback control problems: robust
  stability and performance,''
\newblock {\em IEEE Transactions on Automatic Control}, vol. 41, no. 7, pp.
  997--1003, 1996.

\bibitem{cvx}
M.~Grant and S.~P. Boyd,
\newblock ``{CVX}: Matlab software for disciplined convex programming, version
  2.0 beta,'' {http://cvxr.com/cvx}, Sept. 2012.

\end{thebibliography}
\end{document}